\documentclass[10pt]{article}
\usepackage{amssymb,amsmath,epsfig}
\begin{document}

\title{\bf{Dynamical System Analysis of Interacting Hessence Dark Energy in $f(T)$
Gravity}}

\author{\bf{Jyotirmay Das Mandal}\thanks{jdm2015@gmail.com} ~ and~
\bf{Ujjal Debnath}\thanks{ujjaldebnath@gmail.com}\\
Department of Mathematics, Indian Institute of Engineering
Science\\ and Technology, Shibpur, Howrah-711 103, India.\\}
\date{}
\maketitle

\begin{abstract}
In this work, we have carried on dynamical system analysis of
hessence field coupling with dark matter in $f(T)$ gravity. We
have analysed the critical points due to autonomous system. The
resulting autonomous system is non-linear. So, we have approached
via the the theory of non-linear dynamical system. We have noticed
very few papers are devoted to this kind of study. Maximum works
in literature are done treating the dynamical system as done in
linear dynamical analysis, which are unable to predict correct
evolution. Our work is totally different from those kind of works.
We have used theory of non-linear dynamical system theory,
developed till date, in our analysis. This approach gives totally
different stable solutions, in contrast what the linear analysis
would have predicted. We have discussed the stability analysis in
details due to exponential potential through computational method
in tabular form and analyzed the evolution of the universe. Some
plots are drawn to investigate the behaviour of the system (this
plotting technique is different from usual phase plot and devised
by us). Interestingly, the analysis shows the universe may
resemble the `cosmological constant' like evolution (i.e.,
$\Lambda$CDM model is a subset of the solution set). Also, all the
fixed points of our model are able to avoid Big rip singularity.
\end{abstract}

\section{\textit{Introduction}}

High end cosmological observations of the Supernova of type Ia (SN
Ia), WMAP, etc.,
\cite{1,1a,2,2a,3,3a,3b,3c,3d,3e,3f,4,4a,4b,4c,4d,4e,5,6} suggest
the fact that the universe may be accelerating lately again after
the early phase. Many theories are formulated to explain this late
time acceleration. However, these theories can be divided mainly
in two categories fulfilling the criteria of a homogeneous and
isotropic universe. First kind of theory (better to known as
`Standard model' or $\Lambda$CDM model) assumes a fluid of
negative pressure named as `dark energy' (DE). The name arises
from the fact the exact origin of this energy is still unexplained
in theoretical set up. Observations, anyway, indicate nearly
70$\%$ of the universe may be occupied by this kind of energy.
Dust matter (cold dark matter (CDM) and baryon matter) comprises
the rest 30$\%$ and there is negligible radiation. Cosmologists
are inclined to suspect dark energy as the primal cause of the
late acceleration of universe. Theory of dark energy has remained
one of the foremost area of research in cosmology till the
discovery of acceleration of the universe at late times
\cite{7,7a,7b,7c,8,8a}. One could clearly notice from the second
field equation, that the expansion would be accelerated if the
equation of state (EoS) parameter satisfies, $ p/{\rho} {\equiv}
{\omega} < -1/3$. Accordingly, then a priori choice for dark
energy is a time independent positive `cosmological constant'
which relates to the equation of state (EoS) $\omega=-1$. This
gives an universe which is expanding forever at exponential rate.
Anyway, cosmological constant has some severe shortcomings like
fine tuning problem etc (see\cite{7} for a review), some recent
data \cite{m8a-A1n,m8a-A1nn} in some sense, agrees with this
choice. By the way, observations which constrains $\omega$ close
to the value of cosmological constant, of $\omega$ does not
indicate whether $\omega$ changes with time or not. So
theoretically, one could consider $\omega$ as a function of cosmic
time, such as inflationary cosmology (see \cite{A1,B2,C3,D4,O15}
for review). Scalar fields evolve in particle physics quite
naturally. Till date, a large variety of scalar field inflationary
models  are discussed. This theory is active area in literature
nowadays (see \cite{7}). The scalar field which lightly interacts
with gravity is called `quintessence'. Quintessence fields are
first hand choice because this field can lessen fine tuning
problem of cosmological constant to some extent. Needless to say,
some common drawback for quintessence are also there. Observations
point that at current epoch energy density of scalar field and
matter energy density is comparable. But, we know they evolve from
different initial conditions. This discrepancy (known as
`coincidence problem') arises for any scalar field dark energy,
quintessence too suffer from this problem \cite{mO-Wn1}. Of
course, there is resolution of this problem, they are called
`tracking solution' \cite{mO-Wn2}. In the tracking regime, field
value should be of the order of Planck mass. Anyway a general
setback is that we always need to seek for such potentials (see
\cite{mO-Wn3} for related discussion). Eos parameter $\omega$ of
quintessence satisfies $-1{\leq}\omega{\leq}1$. Some current data
indicates that $\omega$ lies in small neighbourhood of
$\omega=-1$. Hence it is technically feasible to relax $\omega$ to
go down the line $\omega<-1$ \cite{mO-Wn4}. There exists another
scalar field with negative kinetic energy term,which can describe
late acceleration. This is named as phantom field, which has Eos
$\omega<-1$ (see details in \cite{7,mO-Wn5}). Phantom field energy
density increases with time. As a result, Hubble factor and
curvature diverges in finite time causing `Big-rip' singularity
(see \cite{mO-Wn6,mO-Wn7,mO-Wn8}). By the way, some specific
choice of potential can avoid this flaw. Present data perhaps
favours a dark energy model with $\omega>-1$ of recent past to
$\omega<-1$ at present time \cite{mO-Wn9}. The line $\omega=-1$ is
known as `phantom' divide. Evidently, neither quintessence nor
phantom field alone can cross the phantom divide. In this
direction, a firsthand choice is to combine both quintessence and
phantom field. This is known in the literature as `quintom' (i.e.,
hybrid of quintessence and phantom) \cite{mO-Wn9}. This can serve
the purpose, but still has some fallacy. A single canonical
complex field is quite natural and useful (like `spintessence'
model \cite{mO-Wn10,mO-Wn11}). However, canonical complex scalar
fields suffer a serious setback, namely the formation of `Q-ball'
(a kind of stable non-topological soliton) \cite{mO-Wn10,mO-Wn11}.\\

To overcome various difficulties with above mentioned models Wei
et al in their paper \cite{W23,W23n} introduced a non-canonical
complex scalar field which plays the role of quintom
\cite{W24,W24n,W24nn}. They name this unique model as `hessence'.
However, hessence is unlike other canonical complex scalar fields
which suffer from the formation of Q-ball. Second kind of theory
modifies the classical general relativity (GR) by higher degree
curvature terms (namely, $f(R)$ theory) \cite{1n,2n,3n} or by
replacing symmetric Levi-Civita connection in GR theory by
antisymmetric Weitzenb$\ddot{o}$ck connection. In other words,
torsion is taken for gravitational interaction instead of
curvature. The resulting theory \cite{4n,5n,6n} (called
`Teleparallel' gravity) was considered initially by Einstein to
unify gravity with electromagnetism in non-Riemannian
Weitzenb\"{o}ck manifold. Later further modification done to
obtain $f(T)$ gravity as in the same vein of $f(R)$ gravity theory
\cite{7n}. Although,the Eos of `cosmological constant'
($\Lambda$CDM model) is well within the various dataset, till now
not a single observation can detect DE or DM, and search for
possible alternative are on their way \cite{mCh7n}. In this regard
alternate gravity theory (like, $f(T)$) really worth discussing.
The work \cite{mC7n} is a nice account in establishing matter stability
of $f(T)$ theory in weak field limit in contrast to $f(R)$ theory.
It is shown that any choice of $f(T)$ can be used.Other reason for
the theoretical advantage for their choice are discussed in the next section.\\

We, in this work have chosen hessence in $f(T)$ gravity. Since,
the system is complex, we have preferred a dynamical analysis. As,
we have mentioned previously hessence field and $f(T)$ theory both
are promising candidates to explain present accelerated phase.
So, we merged them to find if they can highlight present acceleration
more accurately with current dataset. A mixed dynamical system with
tachyon, quintessence and phantom in $f(T)$ theory is considered in
\cite{mC8}. Dynamical systems with quintom  are there in literature also
(see \cite{mC8n,mC8nn} for review). The dynamical system analysis for
normal scalar field model in $f(T)$ gravity has been discussed in ref \cite{Ch}.
But, to the best of our knowledge hessence in $f(T)$ gravity is not considered before.\\

We arrange the paper in following manner. Short sketch of $f(T)$
theory is presented in section 2. Hessence field in $f(T)$ gravity
is introduced to form dynamical system in section 3. Section 4 is
devoted dynamical system analysing and the stability of the system
for hessence dark energy model. The significance of our result is
discussed in section 5 in light of recent data. We concluded the
paper with relevant remarks in section 6. We use
normalized units as $8\pi G={\hbar}=c=1$ in this paper.\\

\section{\textit{A Brief Outline of $f(T)$ gravity: Some Basic Equations}}

In teleparallelism \cite{7n,8n,9n}, $e^\mu_A$ are called the
orthonormal tetrad components $(A=0,1,2,3)$. The index $A$ is used
for each point $x^\mu$ for a tangent space of the manifold, hence
each $e^\mu_A$ represents a tangent vector to the manifold (i.e.,
the so called vierbein). Also the inverse of the vierbein is
obtained from the relation ${e^\mu_A}{e^A_\nu}=\delta^\mu_\nu$.
The metric tensor is given as,
$g_{\mu\nu}=\eta_{AB}{e^A_\mu}{e^A_\nu}$
($\mu,\nu$=0,1,2,3),$\mu,\nu$ are coordinate indices on the
manifold (here, $\eta_{AB}$=diag(1,-1,-1,-1)). Recently, to
explain the acceleration the teleparallel torsion $(T)$ in
Lagrangian density has been modified from linear torsion to some
differentiable function of $T$ \cite{10n,11n} (i.e., $f(T)$)
likewise $f(R)$ theory mentioned earlier. In this new set up of
gravity the field equation is of second order unlike $f(R)$ (which
is fourth order). In $f(T)$ theory of gravitation, corresponding
action reads as,
\begin{equation}\label{2.1}
\mathcal{A}=\frac{1}{2{\kappa^2}}\int{d^4x}[\sqrt{-g}(T+f(T))+\mathcal{L}_m]
\end{equation}
where $T$ is the torsion scalar, $f(T)$ is some differentiable
function of torsion $T$, $\mathcal{L}_m$ is the matter Lagrangian,
$\sqrt{-g}=det(e^A_\mu)$ and $\kappa^2=8\pi G$. The torsion scalar
$T$ mentioned above is defined as,
\begin{equation}\label{2.2}
T={S_\rho}^{\mu\nu}{T^\rho}_{\mu\nu}
\end{equation}
with the components of torsion tensor ${T^\rho}_{\mu\nu}$ of
(\ref{2.2}) is given by,
\begin{equation}\label{2.3}
{T^\rho}_{\mu\nu}={{\Gamma^W}^\lambda}_{\nu\mu}-{{\Gamma^W}^\lambda}_{\mu\nu}
={e^\lambda_A}({\partial_\mu}{e^A_\nu}-{\partial_\nu}{e^A_\mu})
\end{equation}
where,
${{\Gamma^W}^\lambda}_{\nu\mu}={e^\lambda_A}{\partial_\mu}{e^A_\nu}$
is the Weitzenb\"{o}ck connexion. Here, The superpotential
${S_\rho}^{\mu\nu}$ (\ref{2.2}) is defined as bellow,
\begin{equation}\label{2.4}
{S_\rho}^{\mu\nu}=\frac{1}{2}({K^{\mu\nu}}_\rho+{\delta^{\mu}_{\rho}}
{T^{\theta\nu}}_ \theta-{\delta^{\nu}_ {\rho}}{T^{\theta\mu}}_\theta)
\end{equation}
\begin{equation}\label{2.5}
{K^{\mu\nu}}_\rho=(-)\frac{1}{2}({T^{\mu\nu}}_\rho-{T^{\nu\mu}}_\rho-{T_\rho}^{\mu\nu})
\end{equation}
${K^{\mu\nu}}_\rho$ is called as contortion tensor. The contortion
tensor measures the difference between symmetric Levi-Civita
connection and anti-symmetric Weitzenb\"{o}ck conexion. It is easy
to check that the equation of motion reduces to Einstein gravity
if $f(T)=0$. Actually this is the correspondence between
teleparallel and Einsteinian theory \cite{6n}. It is noticed that
$f(T)$ theory can address early acceleration and late evolution of
universe depending on the choice of $f(T)$. For example, power law
or exponential form can't overcome phantom divide \cite{12n}, but
some other choices of $f(T)$ \cite{13n} can cross phantom divide.
The reconstruction of $f(T)$ model \cite{14n,15n}, various
cosmological \cite{16n,17n} and thermodynamical \cite{18n}
analysis, has been reported. It is to interesting to note that
linear $f(T)$ model (i.e., when $\frac{dF}{dT}=$ constant) behaves
as cosmological constant. Anyway, a preferable choice of $f(T)$ is
such that it reduces to General Relativity (GR) when redshift is
large in tune with primordial nucleosynthesis and cosmic microwave
data at early times (i.e., $f/T{\rightarrow}0$ for $a<<1$).
Moreover, in future it should give de-Sitter like state. One such
choice is given in power form as in \cite{19n}, namely
\begin{equation}\label{2.6}
f(T)=\beta {(-T)^n}
\end{equation}
$\beta$ being a constant. In particular, $n=1/2$ gives same
expanding model as the theory referred in \cite{19n,20n}. Current
data needs the bound $`n<<1'$ to permit $f(T)$ as an alternate
gravity theory. The effective DE equation of state varies from
$\omega=-1+n$ of past to $\omega=-1$ in future.\\

Throughout the work we assume flat, homogeneous, isotropic
Friedman-Lema\^{i}tre-Robertson-Walker (FLRW) metric,
\begin{equation*}
ds^2=dt^2-{a^2}(t){\sum}_{i=1}^3{(dx^i)^2}
\end{equation*}
which arises from the vierbein $e^A_\mu=diag(1,a(t),a(t),a(t))$.
Here $a(t)$ is the scale factor as a function of cosmic time
t.Using (\ref{2.3}),(\ref{2.4}),(\ref{2.5}) one gets,
\begin{equation*}
T=S^{\rho\mu\nu}T_{\rho\mu\nu}=-6{H^2}
\end{equation*}
where $H={\dot a(t)}/a(t)$ is the Hubble factor (from here and in
rest of the paper `overdot' will mean the derivative operator
$\frac{d}{dt}$).

\section{\textit{Hessence Dark Energy in $f(T)$ Gravity Theory: Formation of Dynamical Equations }}

Here, we consider a non-canonical complex scalar field
\begin{equation}\label{3.1}
\Phi={\phi_1}+ \textit{i}{\phi_2}
\end{equation}
where, $\textit{i}=\sqrt{-1}$.
with Lagrangian density,
\begin{equation}\label{3.2}
\mathcal{L}_{DE}=\frac{1}{4}[({\partial_\mu}\Phi)^2+({\partial_\mu}\Phi^\ast)^2]
-V(\Phi,\Phi^\ast)
\end{equation}
Clearly the Lagrangian density is identical to the Lagrangian
given by two real scalar fields, which looks like
\begin{equation}\label{3.3}
\mathcal{L}_{DE}=\frac{1}{2}({\partial_\mu}\phi_1)^2-\frac{1}{2}({\partial_\mu}\phi_2)^2
-V(\phi_1,\phi_2)
\end{equation}
where $\phi_1$ and $\phi_2$ are quintessence and phantom fields
respectively. It is noteworthy that, the Lagrangian in (\ref{3.2})
consists of one field, instead of two independent field as in
(\ref{3.3}) of reference \cite{mO-Wn9}. It also differs from
canonical complex scalar field (like `spintessence' in
\cite{mO-Wn10,mO-Wn11}) which has the Lagrangian
\begin{equation}\label{3.4}
\mathcal{L}_{DE}=\frac{1}{2}({\partial_\mu}\Psi^*)({\partial_\mu}\Psi)-V(|\Psi|),
\end{equation}
$|\Psi|$ denote the absolute value of $\Psi$, i.e.,
$|\Psi|^2={\Psi^*}{\Psi}$. However, hessence is unlike canonical
complex scalar fields which suffer from the formation of `Q-ball'
(a kind of stable non-topological soliton). Following Wei et al as
in \cite{W23,W23n}, the energy density $\rho_h$ and pressure $p_h$
of hessence field can be written as,
\begin{equation}\label{3.5}
\rho_h=\frac{1}{2}({\dot{\phi}^2}-\frac{Q^2}{{a^6}{\phi^2}})+V(\phi)
\end{equation}
\begin{equation}\label{3.6}
p_h=\frac{1}{2}({\dot{\phi}^2}-\frac{Q^2}{{a^6}{\phi^2}})-V(\phi)
\end{equation}
where, $Q$ is a constant and denotes the total induced charge in
the physical volume (refer \cite{W23,W23n}). In this paper, we
will consider interaction of hessence field and matter. The matter
is perfect fluid with barotropic equation of state,
\begin{equation}\label{3.7}
p_m={w_m}{\rho_m}{\equiv}(\gamma-1){\rho_m}
\end{equation}
where $\gamma$ is the barotropic index satisfying
$0<\gamma{\leq}2$. Also $p_m$ and $\rho_m$ respectively denotes
the pressure and energy density of matter. In particular
$\gamma=1$ and $\gamma=4/3$ indicate dust matter and radiation
respectively. We suppose hessence and background fluid interacts
through a term $C$. This term $C$ indicates energy transfer
between dark energy and dark matter. Positive $C$ is needed to
solve coincidence problem since positive magnitude of $C$
indicates energy transfer from dark energy to dark matter. Also
$2^{ND}$ law of thermodynamics is also valid with this choice. An
interesting work to settle this problem is reviewed in
\cite{m20n21}. A rigourous dynamical analysis is done there.
Similar approach exists for quintom model too. Various choices of
this interaction term $C$ are used in the literature. Here in view
of dimensional requirement of energy conservation equation and to
make the dynamical system simple, we have taken $C=\delta
\dot{\phi} \rho_m$, where $\delta$ is a real constant of small
magnitude, which may be chosen as positive or negative at will,
such that $C$ remains positive. Also, $\dot{\phi}$ may be positive
or negative according the hessence field $\phi$. So we have,
\begin{equation}\label{3.8}
\dot{\rho}_h+3H(\rho_h+p_h)=-C,
\end{equation}
\begin{equation}\label{3.9}
\dot{\rho}_m+3H(\rho_m+p_m)=C
\end{equation}
preserving the total energy conservation equation
\begin{equation*}
\dot{\rho}_{total}+3H(\rho_{total}+p_{total})=0
\end{equation*}
The modified field equations in $f(T)$ gravity are,
\begin{equation}\label{3.10}
H^2=\frac{1}{(2f_T+1)}\left[\frac{1}{3}(\rho_h+\rho_m)-\frac{f}{6}\right],
\end{equation}
\begin{equation}\label{3.11}
\dot{H}=(-\frac{1}{2})\left[\frac{\rho_h+p_h+\rho_m}{1+{f_T}+2T{f_T}}\right]
\end{equation}
In view of equations (\ref{3.5}) and (\ref{3.8}) we have,
\begin{equation}\label{3.12}
\ddot{\phi}+3H\dot{\phi}+\frac{Q^2}{{a^6}{\phi^2}}+V^{\prime}=-{\delta}{\rho_m}
\end{equation}
Here, `$\prime$' means `$\frac{d}{d\phi}$'. Similarly equations
(\ref{3.7}) and (\ref{3.9}) give,
\begin{equation}\label{3.13}
\dot{\rho}_m+3H{\gamma}{\rho_m}={\delta}{\dot{\phi}}{\rho_m}
\end{equation}
Now, we introduce five auxiliary variables,
\begin{equation}\label{3.14}
x=\frac{\dot{\phi}}{\sqrt{6}H},\quad y=\frac{\sqrt{V}}{\sqrt{3}H},
\quad u=\frac{\sqrt{6}}{\phi},\quad
v=\frac{Q}{\sqrt{6}H\sqrt{{a^3}{\phi}}},\quad
\Omega_m=\frac{\rho_m}{3 {H^2}}
\end{equation}
We form the following autonomous system after some manipulation,
\begin{equation}\label{3.15}
\frac{dx}{dN}=-3x-u{v^2}-{\lambda}{\sqrt{\frac{3}{2}}}{y^2}-{\delta}{\sqrt{\frac{3}{2}}}
{\Omega_m}+\frac{3x}{2}({2x^2}-{2v^2}+\Omega_m)
\end{equation}
\begin{equation}\label{3.16}
\frac{dy}{dN}={\lambda}{\sqrt{\frac{3}{2}}}xy+\frac{3}{2}y({2x^2}-{2v^2}+\Omega_m)
\end{equation}
\begin{equation}\label{3.17}
\frac{du}{dN}=-x{u^2}
\end{equation}
\begin{equation}\label{3.18}
\frac{dv}{dN}=-xuv-3v+\frac{3}{2}v({2x^2}-{2v^2}+\Omega_m)
\end{equation}
\begin{equation}\label{3.19}
\frac{d\Omega_m}{dN}=\Omega_m(-3\gamma-\delta\sqrt{6}x+3({2x^2}-{2v^2}+\Omega_m))
\end{equation}
In above calculations, $N={\int}\frac{\dot{a}}{a}dt=lna$, denotes
the `e-folding' number. We have chosen $N$ as independent
variable. We have taken $f(T)=\beta \sqrt{-T}$ for above
derivation of autonomous system. Also, we have chosen exponential
form of potential i.e., $\frac{V^\prime}{V}=\lambda$ ( where
$\lambda$ is a real constant) for simplicity of the autonomous
system. This kind of choice is standard in literature with coupled
real scalar field \cite{m20n21-21}, complex field (like, hessence
in loop quantum cosmology) in \cite{mC8nn}. The work \cite{Ch}
dealing quintessense, matter in $f(T)$ theory, is also done with
exponential potential. But, to our knowledge hessense, matter in
$f(T)$ theory is not considered before. In view of (\ref{3.14}),
the Friedmann equation (\ref{3.10}) reduces as,
\begin{equation}\label{3.20}
{x^2}+{y^2}-{v^2}+{\Omega_m}=1
\end{equation}
The Raychoudhury equation becomes,
\begin{equation}\label{3.21}
-\frac{\dot{H}}{H^2}=\frac{3}{2}(2{x^2}-2{v^2}+{\Omega_m})
\end{equation}
The density parameter of hessence ($\Omega_h$) dark energy and
background matter ($\Omega_m$) are obtained in the following
forms:,
\begin{equation}\label{3.22}
\Omega_h=\frac{\rho_h}{3{H^2}}={x^2}+{y^2}-{v^2}, {\quad} \Omega_m=\frac{\rho_m}{3{H^2}}=1-({x^2}+{y^2}-{v^2})
\end{equation}
The Eos of hessence $\omega_h$ dark energy and total Eos of the
system $\omega_{total}$ are calculated in the forms:
\begin{equation}\label{3.23}
\omega_h=\frac{p_h}{\rho_h}=\frac{{x^2}-{y^2}-{v^2}}{{x^2}+{y^2}-{v^2}}, {\quad}
\omega_{total}=\frac{p_h+p_m}{\rho_h+\rho_m}={x^2}-{y^2}-{v^2}+(\gamma-1){\Omega_m}
\end{equation}
Also, the deceleration parameter $q$ can be expressed as
\begin{equation}\label{3.24}
q=-1-\frac{\dot{H}}{H^2}=-1+\frac{3}{2}(2{x^2}-2{v^2}+{\Omega_m})
\end{equation}

\section{\textit{Fixed Points and Stability Analysis of the Autonomous System}}

\subsection{\textit{Fixed Points with Exponential Potential}}

We have made the choice of exponential form of potential i.e.,
$\frac{V^\prime}{V}=\lambda$ (where $\lambda$ is a real constant).
The fixed points $P_i$, the co-ordinates of $P_i$ i.e.,
(${x_c}$,${y_c}$,${u_c}$,${v_c}$,${{\Omega_m}_c}$) are given in
Table $\bf{1}$ with relevant parameters and existence
condition(s).

\begin{table}[h]
\begin{tabular}{|l|l|l|l|l|l|l|l|}
\hline
$P_i$ & ${x_c}$ , ${y_c}$ , ${u_c}$ , ${v_c}$ , ${{\Omega_m}_c}$ & ${\Omega_m}$ & ${\omega_h}$ & ${\omega_{total}}$ & ${\Omega_h}$ & q & Existence  \\
\hline
$P_1$ & 1 , 0 , 0 , 0 , 0 & 0 & 1 & 1 & 1 & 2 & always \\
\hline
$P_2$ & -1 , 0 , 0 , 0 , 0 & 0 & 1 & 1 & 1 & 2 & always \\
\hline
$P_{3\pm}$ & ${\pm}1$ or $\frac{6-3\gamma}{\delta\sqrt{6}}$ , 0 , 0 , 0 , 0 & 0 & 1 & 1 & 1 & 2 & $\frac{6-3\gamma}{\delta\sqrt{6}}={\pm}1$\\
\hline
$P_4$ & $-\sqrt{\frac{2}{3}}\delta$ , 0 , 0 , 0 , $\frac{6-3\gamma+2\delta^2}{3}$ & $\frac{6-3\gamma+2\delta^2}{3}$ & 1 & $\gamma(1-\frac{2{\delta^2}}{3})$ & $\frac{2{\delta^2}}{3}$ & $\frac{1}{2}+{\delta^2}$ & $\frac{2\delta^2}{3}+\frac{6-3\gamma+2\delta^2}{3}=1$\\

 &  &  &  & $+ \frac{4{\delta^2}}{3}-1$ &  &  & \\
\hline
$P_5$ & $x=\frac{6-3\gamma}{\delta\sqrt{6}}$ , 0 , 0 , $\sqrt{{x^2}-1}$ , 0 & 0 & 1 & 1 & 1& 2 & $6{\delta^2}{\leq}{(6-3\gamma)^2}$\\
\hline
$P_6 $& $x=\frac{6-3\gamma}{\delta\sqrt{6}}$ , 0 , 0 , $-\sqrt{{x^2}-1}$ , 0 & 0 & 1 & 1 & 1& 2 & $6{\delta^2}{\leq}{(6-3\gamma)^2}$\\
\hline
$P_7$ & 0 , 1 , any value , 0 , 0 & 0 & -1 & -1 & 1 & -1 & $\gamma=0$\\
\hline
$P_8$ & 0 , 1 , any value , 0 , 0 & 0 & -1 & -1 & 1 & -1 & $\gamma=0$\\
\hline
$P_9$ & $-\frac{\lambda}{\sqrt{6}}$ , $\sqrt{1-\frac{\lambda^2}{6}}$ , 0 , 0 , 0 & 0 & $-1 + \frac{\lambda^2}{3}$ & $-1 + \frac{\lambda^2}{3}$ & 1 & $-1 + \frac{\lambda^2}{2}$ &${\lambda^2}{\leq}6$\\
\hline
$P_{10}$ & $-\frac{\lambda}{\sqrt{6}}$ , $-\sqrt{1-\frac{\lambda^2}{6}}$ , 0 , 0 , 0 & 0 & $-1 + \frac{\lambda^2}{3}$ & $-1 + \frac{\lambda^2}{3}$ & 1 & $-1 + \frac{\lambda^2}{2}$ &${\lambda^2}{\leq}6$\\
\hline
$P_{11}$ & A , $\sqrt{1-{A^2}-{B^2}}$ , 0 , 0 , B & B & $\frac{-1+2{A^2}+{B^2}}{1-B}$ & $-1+{A^2}+{B^2}$ & 1-B & $-1+\frac{3}{2}B$ &$\delta+\lambda{\neq}0 $\\

 &  &  &  & $+(\gamma-2)B$ &  & $+3{B^2}$ &\\
\hline
$P_{12}$ & A , $-\sqrt{1-{A^2}-{B^2}}$ , 0 , 0 , B & B & $\frac{-1+2{A^2}+{B^2}}{1-B}$ & $-1+{A^2}+{B^2}$ & 1-B & $-1+\frac{3}{2}B$ & $\delta+\lambda{\neq}0 $\\

 &  &  &  & $+(\gamma-2)B$ & & $+3{B^2}$ & \\
\hline
$P_{13}$ & $ -\frac{\sqrt{6}}{\lambda}$ , 0 , 0 , $\sqrt{\frac{6}{\lambda^2}-1}$ , 0 & 0 & 1 & 1 & 1 & 2 & ${\lambda^2}{\leq}6$\\
\hline
$P_{14}$ & $ -\frac{\sqrt{6}}{\lambda}$ , 0 , 0 , $-\sqrt{\frac{6}{\lambda^2}-1}$ , 0 & 0 & 1 & 1 & 1 & 2 & ${\lambda^2}{\leq}6$\\
\hline
$P_{15}$ & $x=-\frac{6}{\lambda}=\frac{6-3\gamma}{\delta \sqrt{6}}$ , 0 , 0 , $\sqrt{x^2-1}$ , 0 & 0  & 1 & 1 & 1 & 2 &$-\frac{6}{\lambda}=\frac{6-3\gamma}{\delta\sqrt{6}}$ \\
\hline
$P_{16}$ & $x=-\frac{6}{\lambda}=\frac{6-3\gamma}{\delta \sqrt{6}}$ , 0 , 0 , $-\sqrt{x^2-1}$ , 0 & 0  & 1 & 1 & 1 & 2 & $-\frac{6}{\lambda}=\frac{6-3\gamma}{\delta\sqrt{6}}$  \\
\hline
\end{tabular}
\caption{Fixed points of the autonomous system of equations
(\ref{3.15})-(\ref{3.19}) and various physical parameters with
existence conditions. Here
$A=-\sqrt{\frac{3}{2}}~\frac{\gamma}{\delta+\lambda}$ and
$B=\frac{6+\lambda \sqrt{6}A-6{A^2}}{9}$.}
\end{table}

From Table$\bf{1}$ we note that,
${\bullet}$ Case $\bf{1}$: Fixed points $P_1,P_2=({\pm1} , 0 , 0 , 0 , 0)$ always exist with the physical parameter ${\Omega_m}=0$, ${\omega_h}=1$, ${\omega_{total}}=1$,  ${\Omega_h}=1$, $q=2$.\\
${\bullet}$ Case $\bf{2}$: Fixed point $P_{3\pm}=({\pm1}$ or $\frac{6-3\gamma}{\delta\sqrt{6}}$ , 0 , 0 , 0 , 0 )  exists under the condition the $\frac{6-3\gamma}{\delta\sqrt{6}}={\pm1}$ with the physical parameter ${\Omega_m}=0$, ${\omega_h}=1$, ${\omega_{total}}=1$,  ${\Omega_h}=1$, $q=2$, i.e., same as $P_1$ and $P_2$.\\
${\bullet}$ Case $\bf{3}$: Fixed point $P_4=(-\sqrt{\frac{2}{3}}\delta$ , 0 , 0 , 0 , $\frac{6-3\gamma+2\delta^2}{3}$)  exists under the condition the $\frac{2\delta^2}{3}+\frac{6-3\gamma+2\delta^2}{3}=1$ with physical parameter ${\Omega_m}=\frac{6-3\gamma+2\delta^2}{3}$, ${\omega_h}=1$, ${\omega_{total}}=-1 + \gamma(1-\frac{2{\delta^2}}{3})+ \frac{4{\delta^2}}{3}$,  ${\Omega_h}=\frac{2{\delta^2}}{3}$, $q=\frac{1}{2}+{\delta^2}$.\\
${\bullet}$ Case $\bf{4}$: Fixed points $P_5,P_6=(x=\frac{6-3\gamma}{\delta\sqrt{6}}$ , 0 , 0 , ${\pm}\sqrt{{x^2}-1}$ , 0)  exist under the condition the $6{\delta^2}{\leq}{(6-3\gamma)^2}$ with physical parameter ${\Omega_m}=0$, ${\omega_h}=1$, ${\omega_{total}}=1$,  ${\Omega_h}=1$, $q=2$.\\
${\bullet}$ Case $\bf{5}$: Fixed points $P_7,P_8=(0 , 1 , any value , 0 , 0)$  exist under the condition the $\gamma=0$ with physical parameter ${\Omega_m}=0$, ${\omega_h}=-1$, ${\omega_{total}}=-1$,  ${\Omega_h}=1$, $q=-1$.\\
${\bullet}$ Case $\bf{6}$: Fixed points $P_9,P_{10}=(-\frac{\lambda}{\sqrt{6}}$ , ${\pm}\sqrt{1-\frac{\lambda^2}{6}}$ , 0 , 0 , 0) exist under the condition the ${\lambda^2}{\leq}6$ with physical parameter ${\Omega_m}=0$, ${\omega_h}=-1+\frac{\lambda^2}{3}$, ${\omega_{total}}=-1+\frac{\lambda^2}{3}$,  ${\Omega_h}=1$, $q=-1+\frac{\lambda^2}{2}$.\\
${\bullet}$ Case $\bf{7}$: Fixed points $P_{11},P_{12}=(A , {\pm}\sqrt{1-{A^2}-{B^2}}$ , 0 , 0 , B) exist under the condition the $\delta+\lambda{\neq}0$ with physical parameter ${\Omega_m}=B$, ${\omega_h}=\frac{-1+2{A^2}+{B^2}}{1-B}$, ${\omega_{total}}=-1+{A^2}+{B^2}+(\gamma-2)B$,  ${\Omega_h}=1-B$, $q=-1+\frac{3}{2}B+3{B^2}$.\\
${\bullet}$ Case $\bf{8}$: Fixed points $P_{13},P_{14}=(-\frac{\sqrt{6}}{\lambda}$ , 0 , 0 ,${\pm}\sqrt{\frac{6}{\lambda^2}-1}$ , 0 )  exist under the condition the ${\lambda^2}{\leq}6$ with physical parameter ${\Omega_m}=0$, ${\omega_h}=1$, ${\omega_{total}}=1$,  ${\Omega_h}=1$, $q=2$.\\
${\bullet}$ Case $\bf{9}$: Fixed points $P_{15},P_{16}=(x=-\frac{6}{\lambda}=\frac{6-3\gamma}{\delta \sqrt{6}}$ , 0 , 0 , ${\pm}\sqrt{x^2-1}$ , 0)  exist under the condition the $-\frac{6}{\lambda}=\frac{6-3\gamma}{\delta\sqrt{6}}$ and $\lambda{\neq}2\delta$ with physical parameter ${\Omega_m}=0$, ${\omega_h}=1$, ${\omega_{total}}=1$,  ${\Omega_h}=1$, $q=2$.\\

\subsection{\textit{Stabilitity of the Fixed Points}}

Dynamical analysis is a powerful technique to study cosmological
evolution, where exact solution could not be found due to
complicated system. This can be done without any information of
specific initial conditions.The dynamical systems mostly
encountered in cosmological system are non linear system of
differential equations (DE). Here the dynamical system is also non
linear.Very few works in literature is devoted to analyse non
linear dynamical system. But, we used the methods developed till
now,\cite{21}. Also we devised some method (as in the plotting of
the dynamical evolution, use of normally hyperbolic fixed points).
We now analyse stability of the fixed points. In this regard, we
find the eigenvalues of the linear perturbation matrix of the
dynamical system (\ref{3.15})-(\ref{3.19}). Due to the Friedmann
equation (\ref{3.20}) we have four independent perturbed equation.
The eigenvalues of the $4{\times}4$ linear perturbation matrix
corresponding each fixed point $P_i$ are given in Table$\bf{2}$.
Before further discussion we state some basics from non linear
system of differential equation (DE) \cite{21}. If the real part
of each eigenvalue is non-zero, then the fixed point is called
hyperbolic fixed point (otherwise, it is called non hyperbolic).
Let us write a non linear system of DE in $R^n$ (the $n$
dimensional Euclidean plane) as,
\begin{equation}\label{4.1}
\mathbf{\dot{x}=f(x)}
\end{equation}
where, $f:E $${\rightarrow}{R^n}$ is derivable and $E$ is an open
set in $R$. For non linear system the DE can not be written in
matrix form as done in linear system. Near hyperbolic fixed point,
although a non-linear dynamical system could be linearised and
stability of the fixed point is found by Hartman-Grobman theorem.
As we can see from the following, Let ${x_c}$ be a fixed point and
$\zeta(t)$ be the perturbation from ${x_c}$ i.e.,
$\zeta(t)=x-{x_c}$, i.e., $x={x_c}+\zeta(t)$. We find the time
evolution of $\zeta(t)$ for (\ref{4.1})as,

\begin{equation}\label{4.2}
\mathbf{\dot{\zeta}=\frac{d}{dt}(x-{x_c})=\dot{x}=f(x)=f({x_c}+\zeta)}
\end{equation}

Since, $f$ is assumed derivable, we use the Taylor expansion of
$f$ to get,

\begin{equation}\label{4.3}
\mathbf{f({x_c}+\zeta)=f(x_c)+\zeta Df(x_c)+...}
\end{equation}

$Df(x)=\frac{\partial{f_i}}{\partial{x_j}} , i,j=1,2,...,n$, as
$\zeta$ is very small higher order terms are neglected above. As,
$f(x_c)=0$, (\ref{4.2}) reduces to,
\begin{equation}\label{4.3}
\mathbf{\dot{\zeta}=\zeta Df(x_c)}
\end{equation}
This is called the linearization of the DE near a fixed point.
Stability of the fixed point $x_c$ is inferred from the sign of
eigenvalues of Jacobian matrix $Df(x_c)$. If the fixed point is
hyperbolic, then stability is concluded from Hartman Grobman theorem, which states,\\
Theorem(Hartman Grobman):Let the non linear DE (\ref{4.1}) in
$R^n$ where, $f$ is derivable with flow $\phi_t$.If $x_c$ is a
hyperbolic fixed point , then there exists a neighbourhood N of
$x_c$ on which $\phi_t$ is homeomorphic to the flow of
linearization of the DE near $x_c$.\\ But for non hyperbolic fixed
point this cannot be done and the study of stability becomes hard
due to lack of theoretical set up. If at least one eigenvalue
corresponding the fixed point is zero, then it is termed as non
hyperbolic. For this case, we can not find out stability near the
fixed point. Consequently, we have to resort to other techniques
like numerical solution of the system near fixed point and to
study asymptotic behaviour with the help of plot of the solution,
as is done in this work (details are described later). However, we
can find the dimension of stable manifold (if exists) with the
help of centre manifold theorem. There are a separate class of of
important non hyperbolic fixed points known as normally hyperbolic
fixed points,which are rarely considered in literature (see,
\cite{22}). As some fixed points encountered in our work are of
this kind, we state the basics here. We are also interested in non
isolated normally hyperbolic fixed points of a given DE (for
example a curve of fixed points, such a set is called equilibrium
set).If an equilibrium set has only one zero eigenvalue at each
point and all other eigenvalue has non-zero real part then the
equilibrium set is called normally hyperbolic. The stability of
normally hyperbolic fixed point is deduced from invariant manifold theorem, which states,\\
Theorem (Invariant manifold):Let $x=x_c$ be a fixed point of the DE $\dot{x}=f(x)$ on $R^n$ and let
$E^s$, $E^u$ and $E^c$ denote the stable ,unstable and centre subspaces of the linearization
of the DE at $x_c$.Then there exists,\\
a stable manifold $W^s$ tangent to $E^s$,\\
an unstable manifold $W^s$ tangent to $E^u$,and\\
a centre manifold $W^c$ tangent to $E^c$ at $x_c$.In other words,
the stability depends on the sign of remaining eigenvalues. If the
sign of remaining eigenvalues are negative, then the fixed point
is stable, otherwise unstable. Table $\mathbf{2}$ shows the
eigenvalues corresponding the fixed points given in Table
$\mathbf{1}$ and existence for hyperbolic, non-hyperbolic or
normally hyperbolic fixed points with the nature of stability (if
any).

\begin{table}
\begin{tabular}{|l|l|l|}
\hline
$P_i$&Eigenvalues&{Nature of Stability (if exists \dag)}\\
\hline
$P_1$ & 0,0,$3+\delta\sqrt{6},3+\sqrt{\frac{3}{2}}\lambda$ & 2D stable manifold \\
$P_2$ & 0,0,$3-\delta\sqrt{6},3-\sqrt{\frac{3}{2}}\lambda$ & 2D stable manifold \\
$P_{3\pm}$ & 0,0,$3{\pm}\delta\sqrt{6},3{\pm}\sqrt{\frac{3}{2}}\lambda$ & 2D stable manifold\\
$P_4$ & $0,-\frac{3}{2}+{\delta^2},-\frac{3}{2}+{\delta^2},\frac{3}{2}+{\delta^2}
-{\delta}{\lambda}$ & 3D stable manifold\\
$P_5$ & 0,0,$3+\sqrt{6}x\delta,3+\sqrt{\frac{3}{2}}x\lambda $ & 2D stable manifold\\
$P_6$ & 0,0,$3+\sqrt{6}x\delta,3+\sqrt{\frac{3}{2}}x\lambda $ & 2D stable manifold\\
$P_7$ & -3,0,$-3-\sqrt{3(\delta\lambda-{\lambda^2})},-3+\sqrt{3(\delta\lambda
-{\lambda^2})}$ & stable\\
$P_8$ & -3,0,$-3-\sqrt{3(\delta\lambda-{\lambda^2})}$,$-3+\sqrt{3(\delta\lambda
-{\lambda^2})}$ & stable\\
$P_9$ & 0,$-3+\frac{\lambda^2}{2},-3+\frac{\lambda^2}{2},-3-\delta\lambda+{\lambda^2}$                   & 3D stable manifold\\
$P_{10}$ & 0,$-3+\frac{\lambda^2}{2},-3+\frac{\lambda^2}{2},-3-\delta\lambda +{\lambda^2}$ & 3D stable manifold\\
$P_{11}$ & 0,$-3+3{a^2}+3\frac{b^2}{2},\frac{1}{4}(D-\sqrt{\Delta})$
,$\frac{1}{4}(D+\sqrt{\Delta})$  & 3D stable manifold\\
$P_{12}$ & 0,$-3+3{a^2}+3\frac{b^2}{2},\frac{1}{4}(D-\sqrt{\Delta})$
,$\frac{1}{4}(D+\sqrt{\Delta})$  & 3D stable manifold\\
$P_{13}$ & 0,0,0,$-3\frac{(2\delta\lambda-{\lambda^2})}{\lambda^2}$ & 1D stable manifold\\
$P_{14}$ & 0,0,0,$-3\frac{(2 \delta\lambda-{\lambda^2})}{\lambda^2}$ & 1D stable manifold\\
$P_{15}$ & 0,0,$3 + \sqrt{6}x\delta,3+\sqrt{\frac{3}{2}}x\lambda$ & 2D stable manifold\\
$P_{16}$ & 0,0,$3 + \sqrt{6}x\delta,3+\sqrt{\frac{3}{2}}x\lambda$ & 2D stable manifold\\
\hline
\end{tabular}
\caption{Eigenvalues of the fixed points of the autonomous system
of equations (\ref{3.15})-(\ref{3.19}) and the nature of stability
(if any)} where $D=-12 + 24{a^2} +
12{b^2}+2\sqrt{6}a\delta+\sqrt{6}a\lambda$ and
$\Delta=-144{a^2}+144{a^4}+144{a^2}{b^2}+36{b^4}+48\sqrt{6}a\delta-48\sqrt{6}{a^3}
\delta-72\sqrt{6}a{b^2}\delta+24{a^2}{\delta^2}-72\sqrt{6}a\lambda+72\sqrt{6}{a^3}\lambda+
84\sqrt{6}a{b^2}\lambda+48\delta\lambda-72{a^2}\delta\lambda-48{b^2}\delta\lambda-
48{\lambda^2} +54{a^2}{\lambda^2}+48{b^2}{\lambda^2}$.\\
{\dag}Nature of Stability is discussed in details.
\end{table}

We see from Table$\bf{2}$ that each fixed point $P_i$ is non
hyperbolic, except $P_7$ and $P_8$ \textbf{(which are normally
hyperbolic)}. So we cannot use linear stability analysis.Hence,we
have utilised the following scheme to infer the stability of non
hyperbolic fixed points.We find the numerical solutions of the
system of differential equations (\ref{3.15})-(\ref{3.19}). Then,
we have investigated the variation of the dynamical variables
$x,y,u,v,$$\Omega_m$ against e-folding N, which in turn gives the
variation against time $t$ through graphs in the neighbourhood of
each fixed points and notice if the dynamical variables
asymptotically converges to any of the fixed points. In that case
we can say the fixed point is stable (otherwise, unstable). This
method are used nowadays in absence of proper mathematical
analysis of non linear dynamical system. But, we must remember the
method is not full proof. Since, we have to consider the
neighbourhood of N as large as possible (i.e.,
$|N|{\rightarrow}{\infty}$). Because a small perturbation can lead
to unstability. The graphs corresponding to each fixed point are
given and analysed below. We consider the fixed points one by one.

We note from figure $\mathbf{1}$ that $P_1$ is not a stable fixed point.
Similar is the case of $P_2$, as is evident from figure $\mathbf{2}$.
We note that if $\lambda{\leq}-\sqrt{6}$ and $\delta{\leq}-\sqrt{\frac{3}{2}}$
(or, $\lambda{\geq}\sqrt{6}$ and $\delta{\geq}\sqrt{\frac{3}{2}}$)
(equality should occur in one of them), then $P_1$ (or, $P_2$) may
admit 2 dimensional stable manifold corresponding the two negative
eigenvalues with Eos of hessence and total Eos being 1, and universe decelerates.\\
We note that $P_{3\pm}$ bears same feature like $P_1$ and $P_2$. So, none of
$P_1$ ,$ P_2$ and $P_3$ describes the current phase of universe.
The points bear no physical significance.\\

\begin{figure}\centering
\epsfig{file=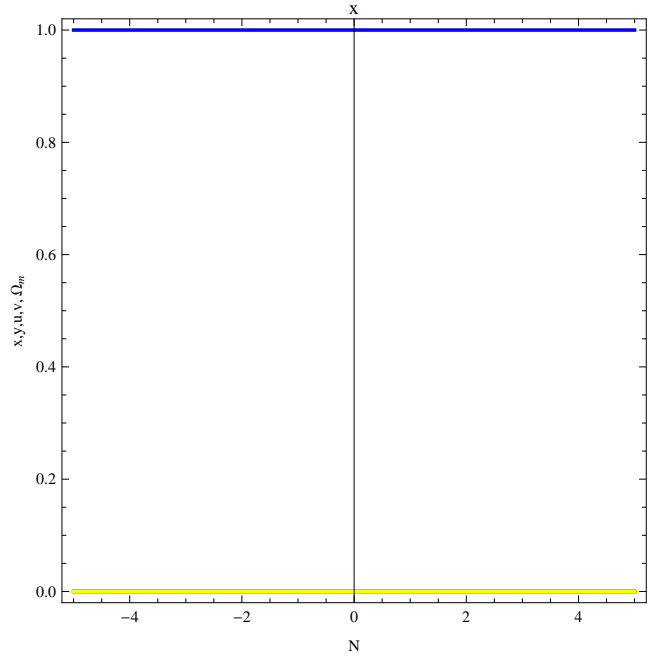,width=.50\linewidth}
\caption{Plot of (1) variations of x (blue),y (green),u
(orange),v (red),$\Omega_m$ (yellow) versus N near $P_1$, for $\gamma=1$ , $\delta=0.5$ and $\lambda=-0.5.$.The position corresponding N=0 is the fixed point
under consideration.}
\end{figure}
\begin{figure}\centering
\epsfig{file=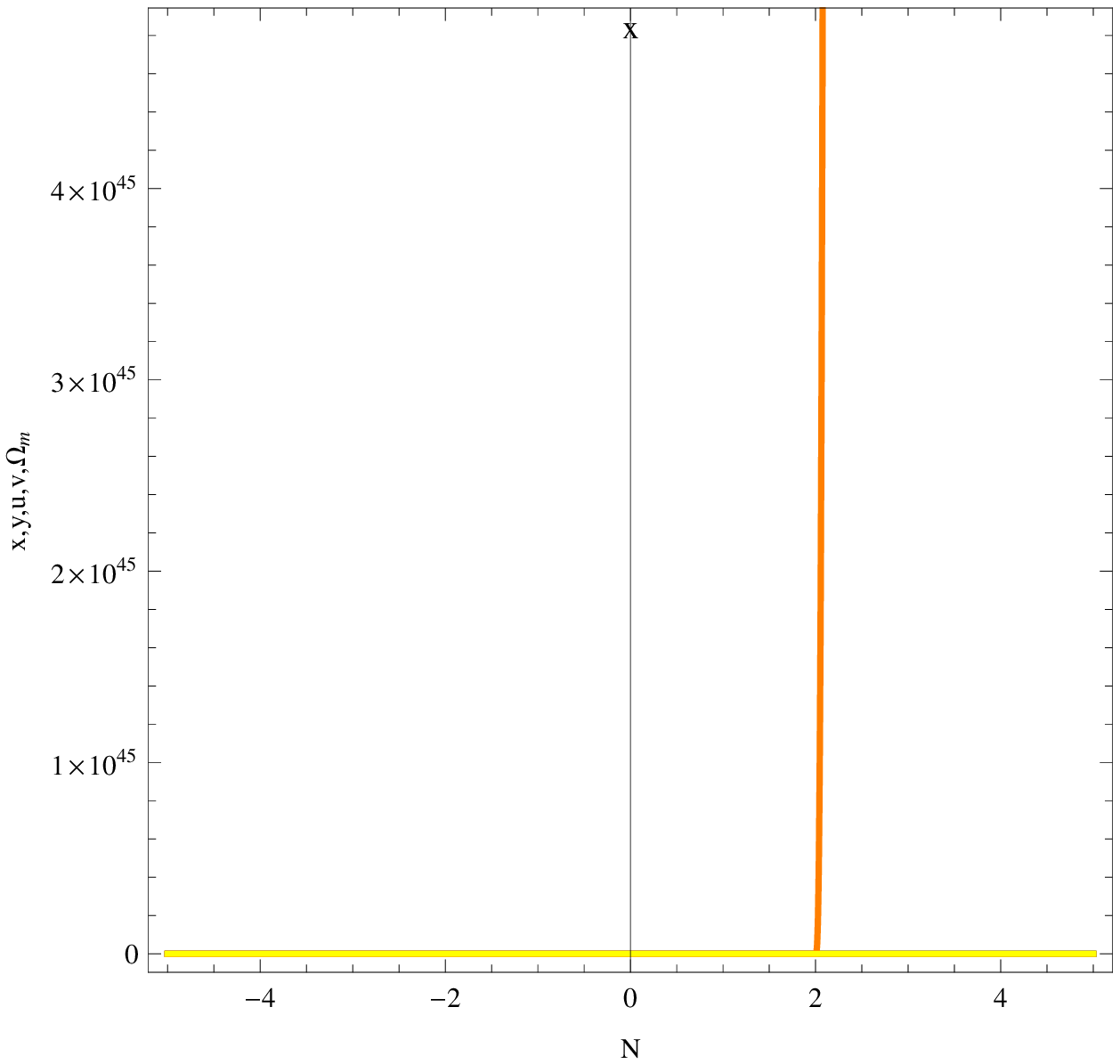,width=.50\linewidth}
\caption{Plot of(2) variations of x (blue),y (green),u (orange),v (red),$\Omega_m$
(yellow) versus N near $P_2$, for $\gamma=1$ , $\delta=0.5$ and
$\lambda=-0.5.$ The position corresponding N=0 is the fixed point
under consideration.}
\end{figure}
\begin{figure}\centering
\epsfig{file=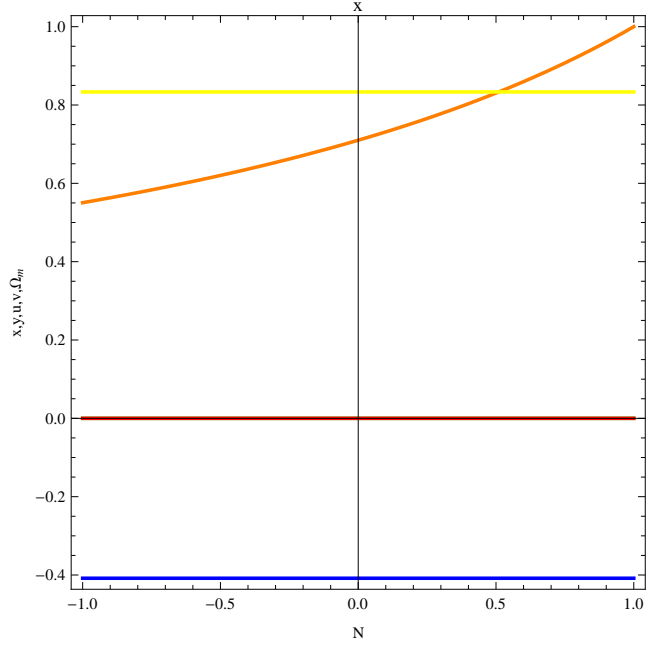,width=.50\linewidth}
\caption{Plot of (3) variations of x (blue),y (green),u (orange),v (red),$\Omega_m$ (yellow) versus N near $P_4$, for $\gamma=4/3$ , $\delta=0.5$ and $\lambda=-0.5.$ The position corresponding N=0 is the fixed point under consideration.}
\end{figure}
If, ${\delta^2}{\leq}\frac{3}{2}$ and ${\delta^2}-\delta\lambda{\leq}-\frac{3}{2}$ (equality should occur in one of them) $P_4$ may admit 2 dimensional stable manifold corresponding the two negative eigenvalues with Eos of hessence is 1 and total Eos is $-1 + \gamma(1-\frac{2{\delta^2}}{3})+ \frac{4{\delta^2}}{3}$ and universe decelerates.Here, the plot figure $\mathbf{3}$ indicates that with a small increase of N the solution moves away from $P_4$.This is a unstable fixed point.\\

\begin{figure}\centering
\epsfig{file=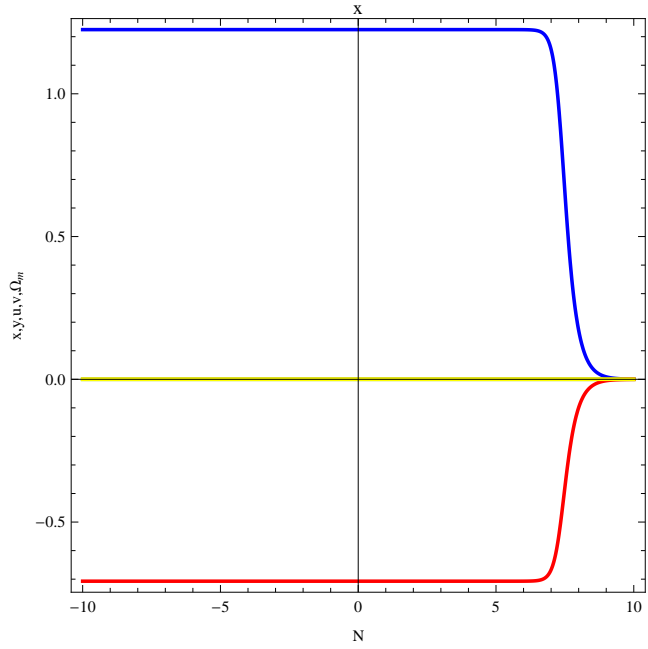,width=.50\linewidth}
\caption{Plot of (4) variations of x (blue),y (green),u
(orange),v (red),$\Omega_m$ (yellow) versus N near $P_5$,for $\gamma=1$ , $\delta=1$ and
$\lambda=-0.5.$ The position corresponding N=0 is the fixed point
under consideration.}
\end{figure}
\begin{figure}\centering
\epsfig{file=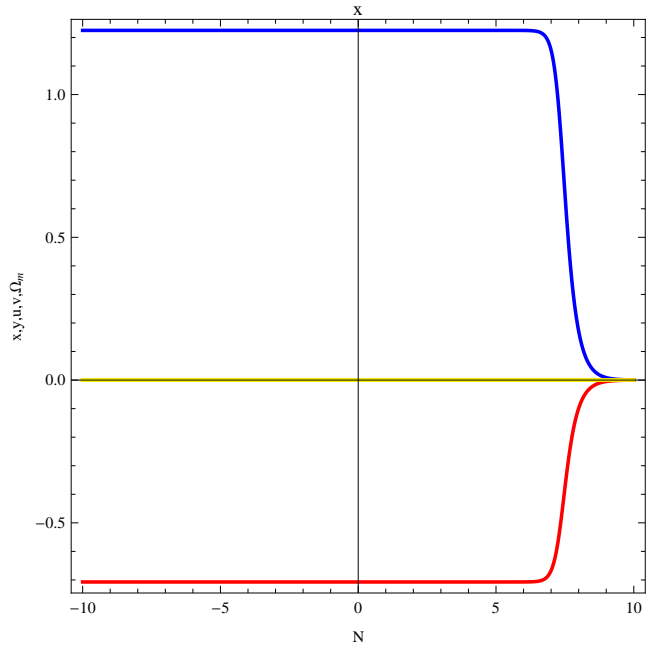,width=.50\linewidth}
\caption{Plot of (5) variations of x (blue),y (green),u (orange),v (red),$\Omega_m$
(yellow) versus N near $P_6$, for $\gamma=1$ , $\delta=1$ and
$\lambda=-0.5.$ The position corresponding N=0 is the fixed point
under consideration.}
\end{figure}

We note that for $P_5$ and $P_6$ if, $x\delta{\leq}-\sqrt{\frac{3}{2}}$ and
$x\lambda {\leq}-\sqrt{6}$ (equality should occur in one of them) $P_5$ may
admit 2 dimensional stable manifold corresponding the two negative eigenvalues
and $P_6$ too may admit 2 dimensional stable manifold corresponding the two
negative eigenvalues with Eos of hessence is 1 and total Eos is 1 and universe
decelerates. The figure $\mathbf{4}$ indicates that the three of the variables
(namely $x, v, $$\Omega_m$) are moving away from $P_5$ and intruding in a
neighbourhood of N=10. This may denote the stable manifold corresponding the
negative eigenvalues.However, this point gives the decelerated phase of the universe.
Similar phenomena can be noted from figure $\mathbf{5}$.\\

\begin{figure}\centering
\epsfig{file=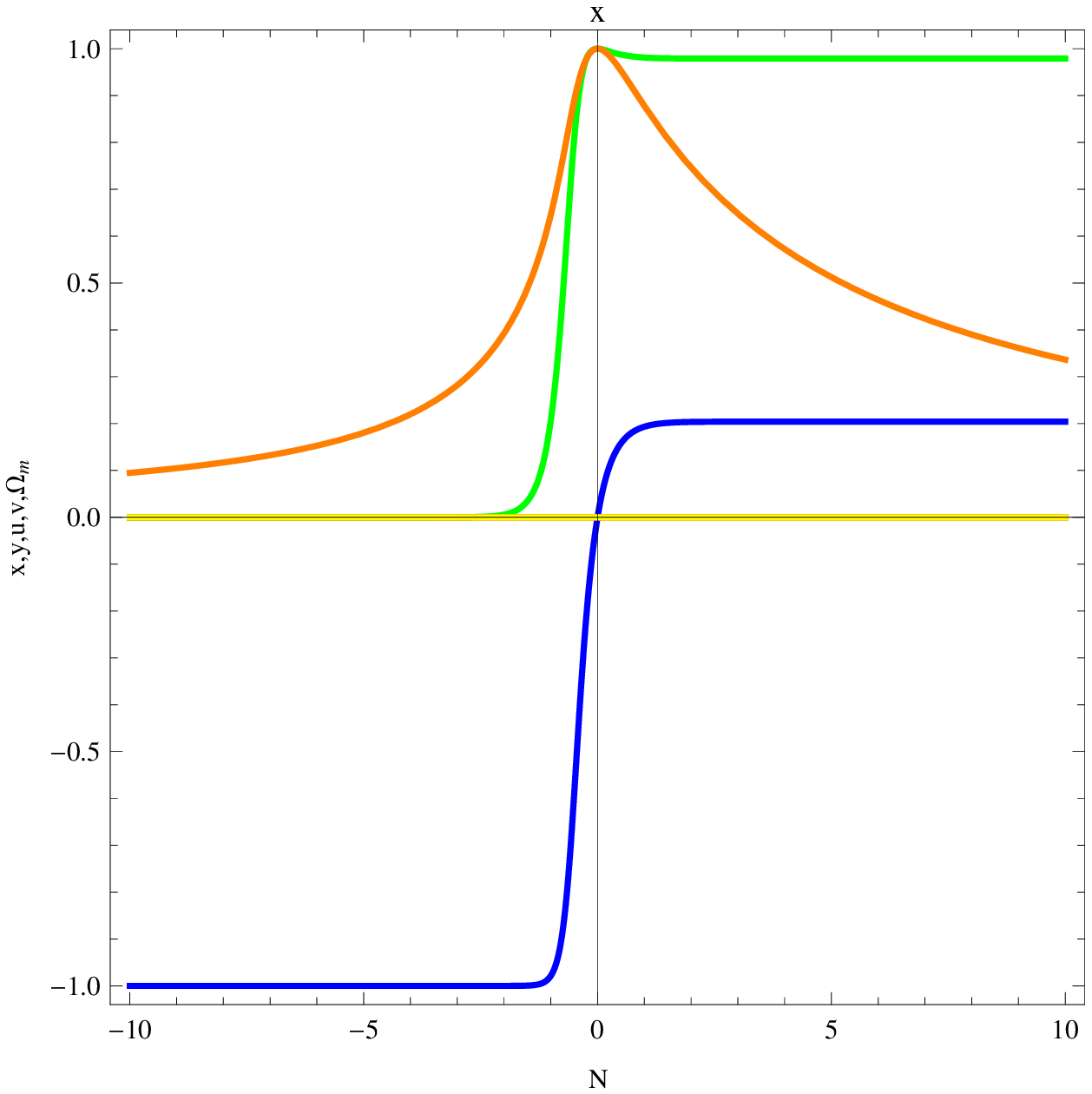,width=.50\linewidth}
\caption{Plot of (6) variations of x (blue),y (green),u
(orange),v (red),$\Omega_m$ (yellow) versus N near $P_7$,for $\gamma=0$ , $\delta=1$ and
$\lambda=-0.5.$ The position corresponding N=0 is the fixed point
under consideration.}
\end{figure}
\begin{figure}\centering
\epsfig{file=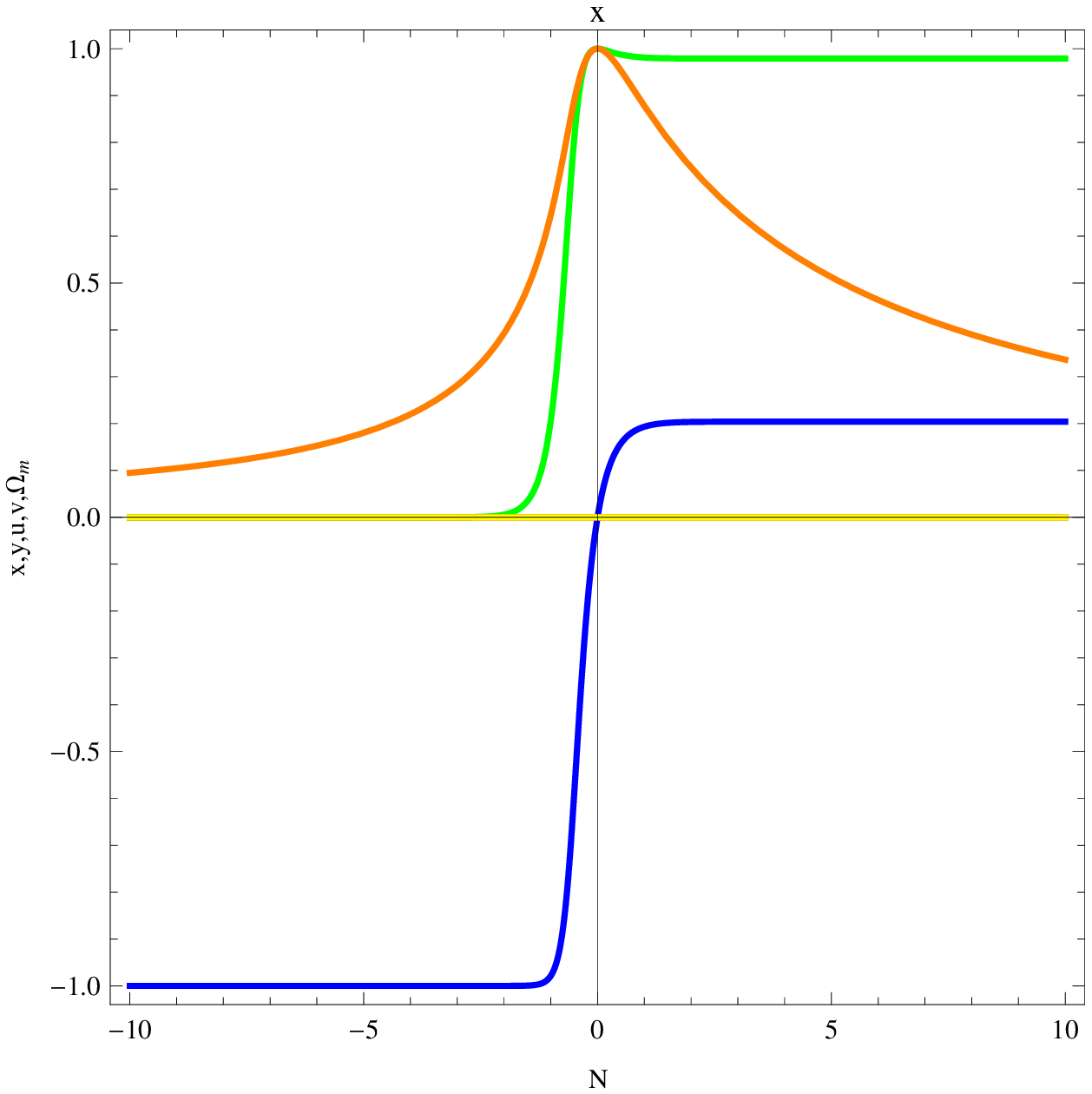,width=.50\linewidth} \caption{Plot of (7)
variations of x (blue),y (green),u (orange),v (red),$\Omega_m$
(yellow) versus N near  $P_8$,for $\gamma=0$ , $\delta=1$ and
$\lambda=-0.5.$ The position corresponding N=0 is the fixed point
under consideration.}
\end{figure}
We note that if $\delta\lambda-{\lambda^2}{\leq}3$ both $P_7$ and
$P_8$ are normally hyperbolic set of fixed points and as the rest
three non-zero eigenvalues are negative they are stable. The set
of fixed points has Eos of hessence is $-1$ and total Eos is also
$-1$ and universe accelerates like `cosmological constant'. We
note clearly from figures $\mathbf{6}$ and $\mathbf{7}$
that all lines from negative and positive values of N (i.e.,
from past and future) are converging towards N=0 (i.e., the set of fixed points).\\

\begin{figure}\centering
\epsfig{file=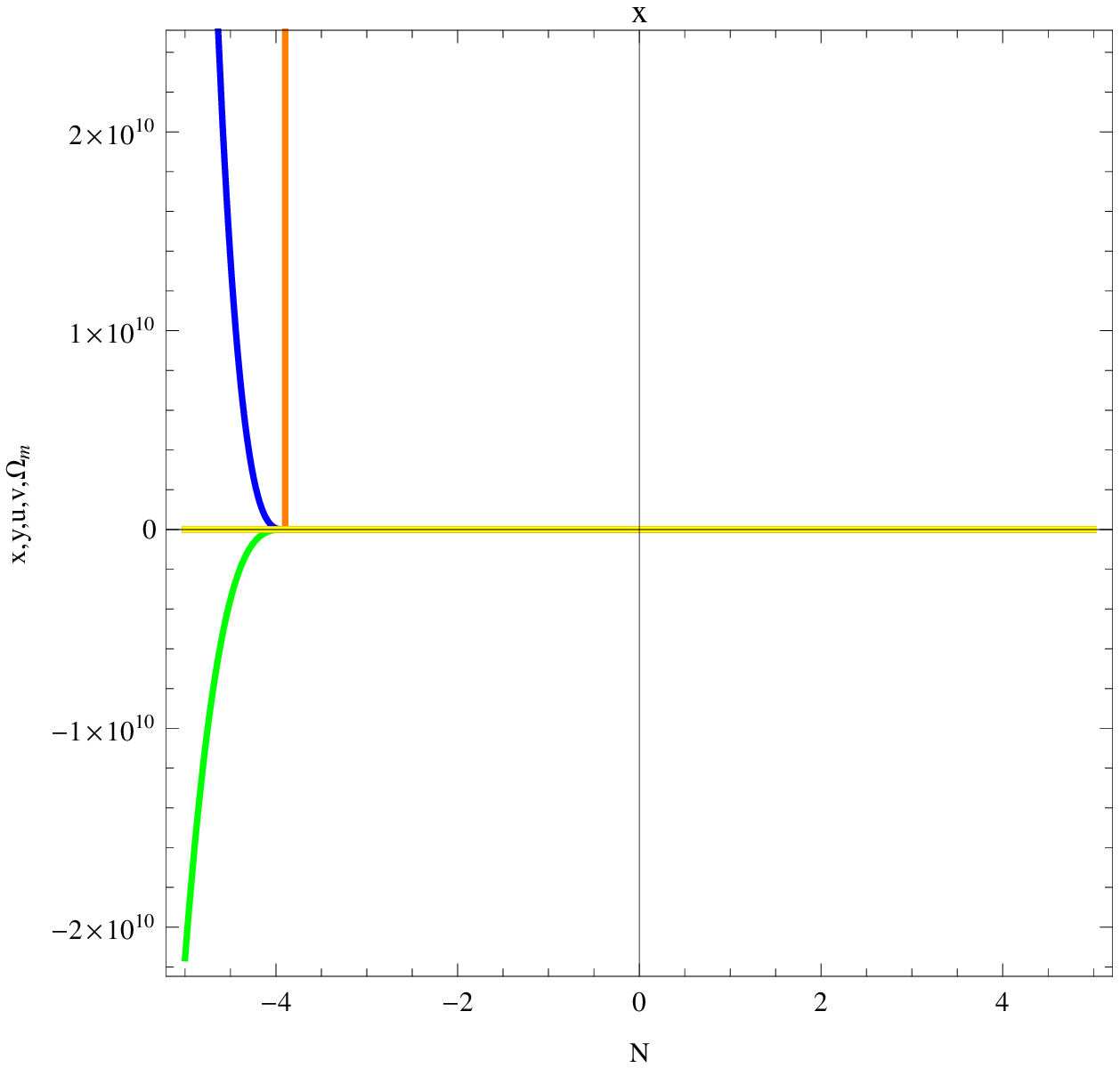,width=.50\linewidth}
\caption{Plot of (8) variations of x (blue),y (green),u
(orange),v (red),$\Omega_m$ (yellow) versus N near $P_9$,for $\gamma=1$ , $\delta=1$ and
$\lambda=-0.5.$ The position corresponding N=0 is the fixed point
under consideration.}
\end{figure}
\begin{figure}\centering
\epsfig{file=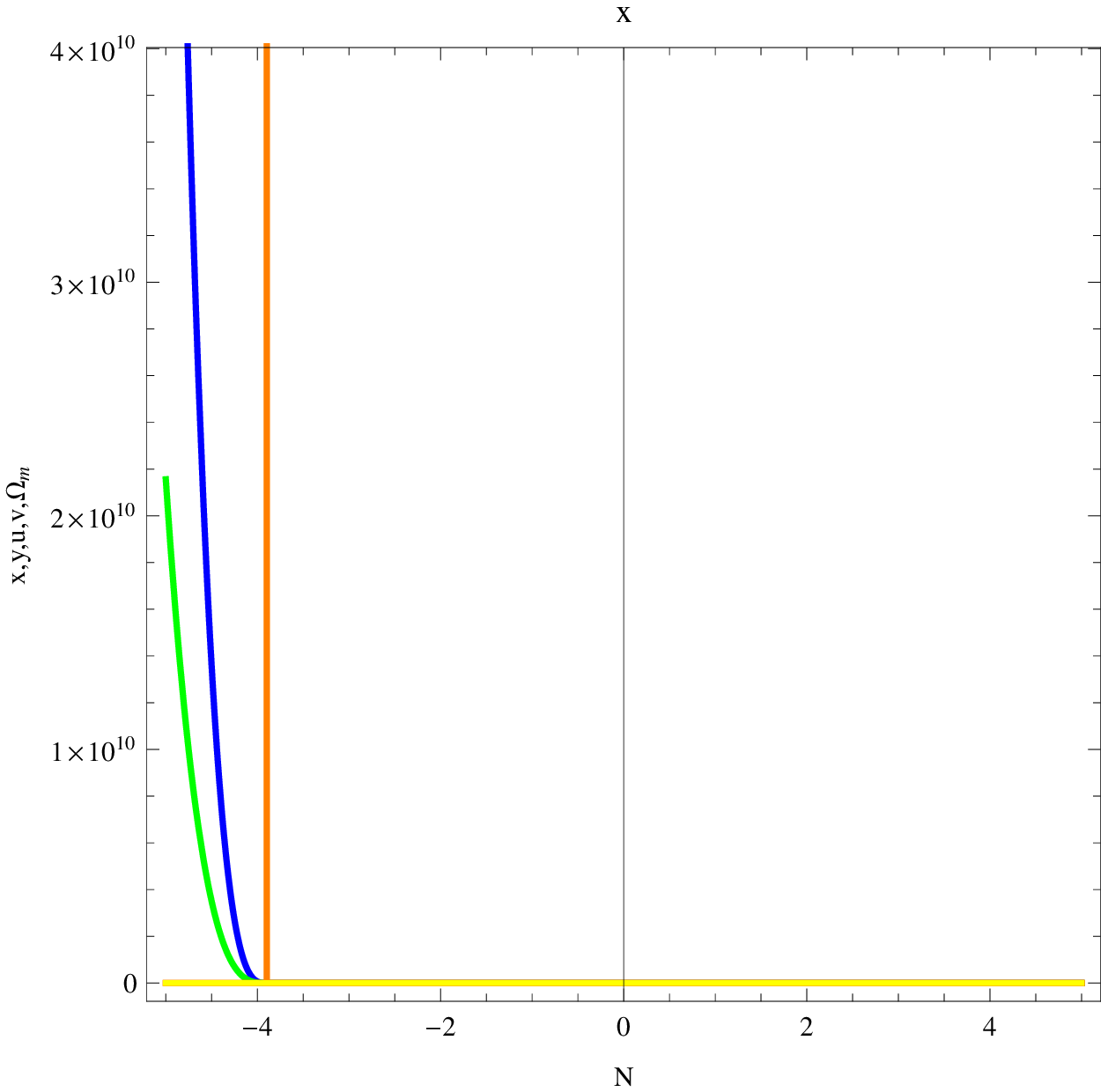,width=.50\linewidth}
\caption{Plot of (9) variations of x (blue),y (green),u (orange),v (red),$\Omega_m$
(yellow) versus N near $P_{10}$,for $\gamma=1$ , $\delta=1$ and
$\lambda=-0.5.$ The position corresponding N=0 is the fixed point
under consideration.}
\end{figure}
We note that if ${\lambda^2}{\leq}6$ and
${\lambda^2}-\delta\lambda{\leq}3$ (equality should occur in one
of them) $P_9$ and $P_{10}$ may admit 3 dimensional stable
manifold corresponding the  negative eigenvalues with Eos of
hessence $-1 + \frac{\lambda^2}{3}$ and total Eos also $-1 +
\frac{\lambda^2}{3}$, (i.e., both Eos are `quintessencelike' if
${\lambda^2}<3$ or `dustlike' if ${\lambda^2}=3$ ). The graphs in
figure $\mathbf{8}$ and $\mathbf{9}$ also supports the fact
corresponding the stable manifolds. In our choice of
$\lambda=-0.5$, Eos of hessence and total Eos, both behaves like `quintessence'.\\

\begin{figure}\centering
\epsfig{file=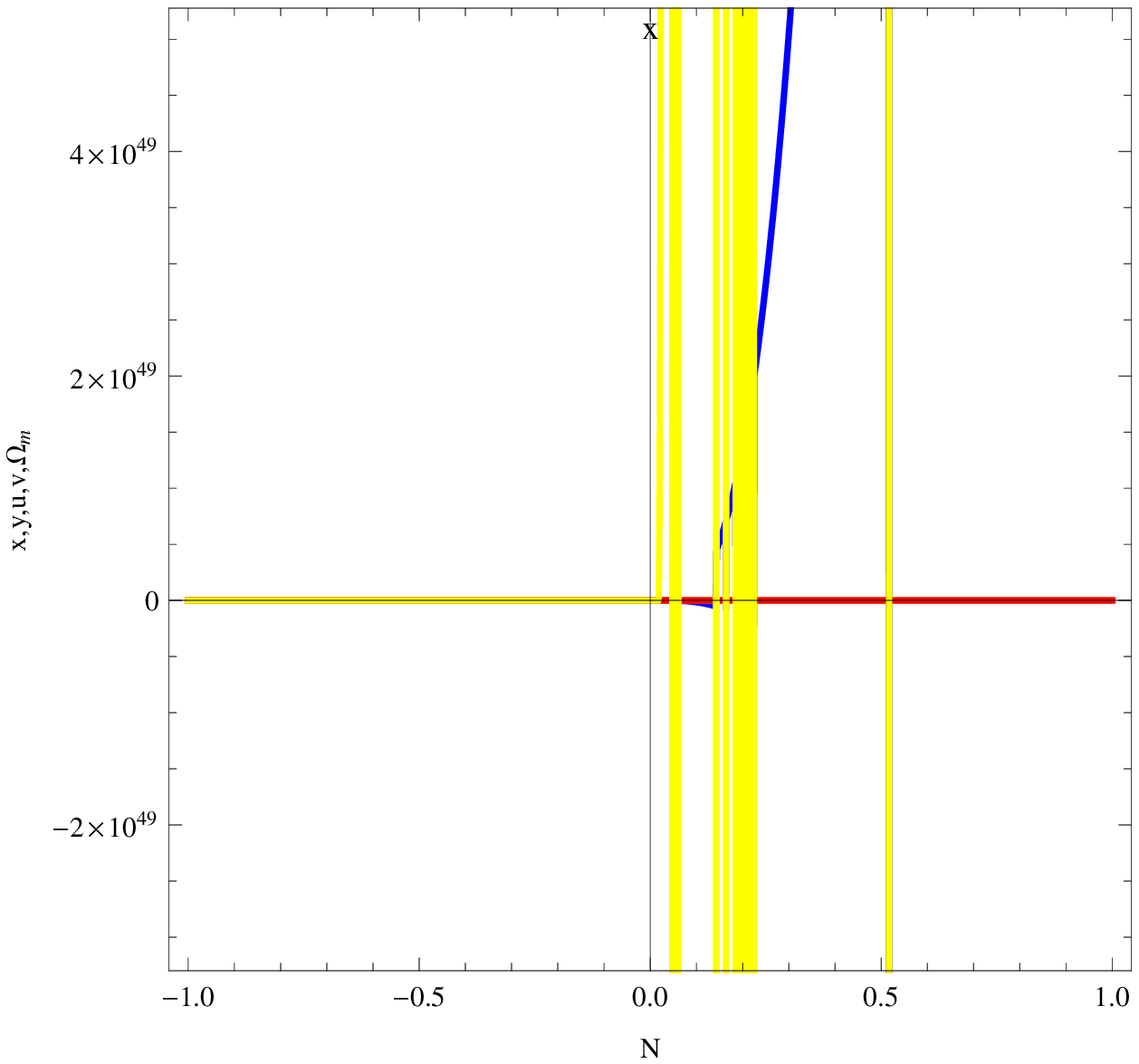,width=.50\linewidth} \caption{Plot of (10)
variations of x (blue),y (green),u (orange),v (red),$\Omega_m$
(yellow) versus N near $P_{11}$,for $\gamma=1$ , $\delta=1$ and
$\lambda=-0.5.$  The position corresponding N=0 is the fixed point
under consideration.}
\end{figure}
\begin{figure}\centering
\epsfig{file=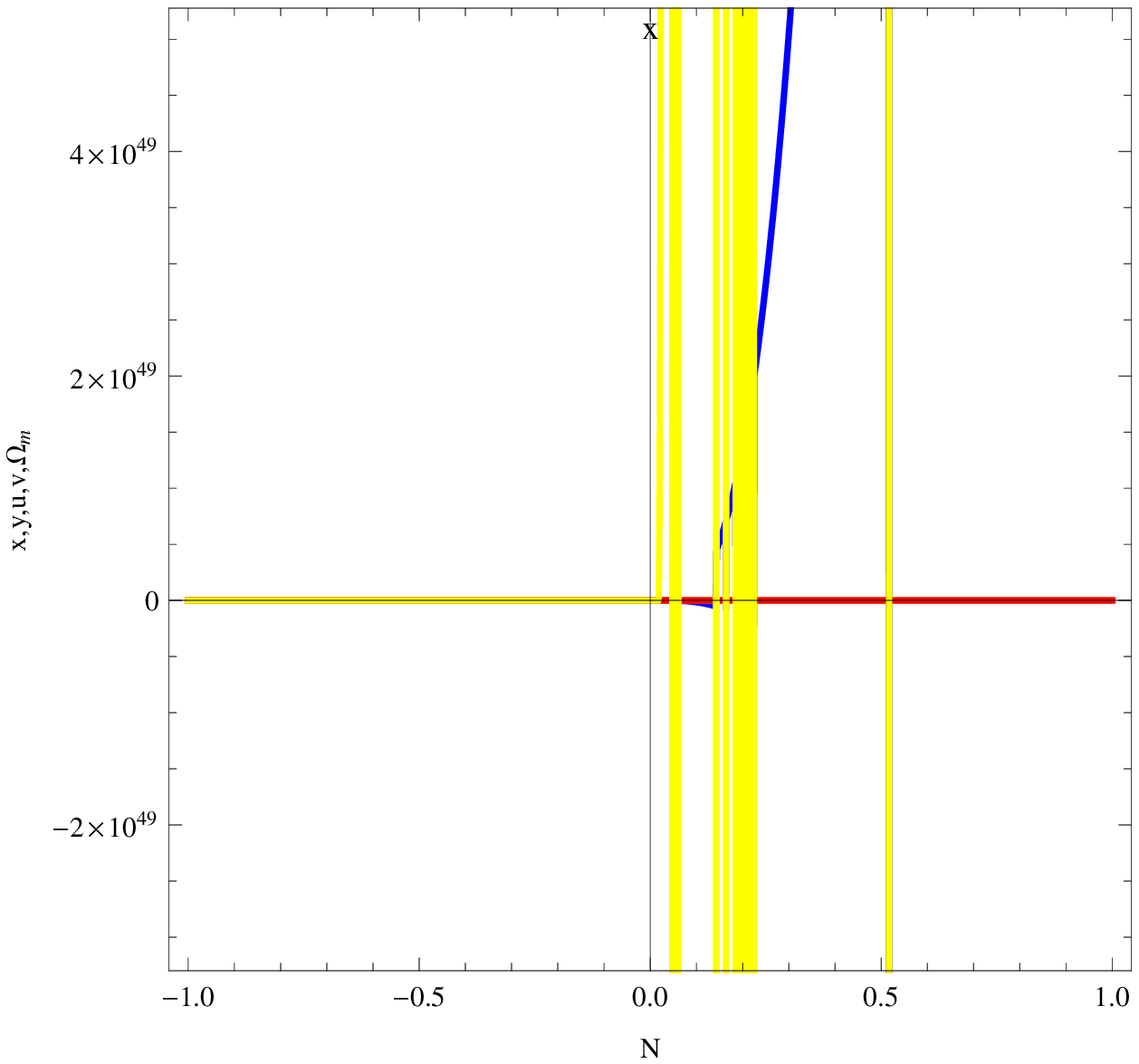,width=.50\linewidth}
\caption{Plot of (11) variations of x (blue),y (green),u (orange),v (red),$\Omega_m$
(yellow) versus N near $P_{12}$,for $\gamma=1$ , $\delta=1$ and
$\lambda=-0.5.$ The position corresponding N=0 is the fixed point
under consideration.}
\end{figure}
We note that if ${a^2}+\frac{b^2}{2}{\leq}1,D{\leq}-\sqrt{\Delta}$
(equality should occur in one of them) $P_{11}$ and $P_{12}$ may admit 3
dimensional stable manifold corresponding the  negative eigenvalues with
Eos of hessence $\frac{-1+2{A^2}+{B^2}}{1-B}$ and total Eos  $-1+{A^2}+{B^2}+(\gamma-2)B$.
We see from figure $\mathbf{10}$ that the system is moving away from the
fixed point $P_{11}$. Similar phenomena happens for fixed point $P_{12}$
as seen from figure $\mathbf{11}$.\\

\begin{figure}\centering
\epsfig{file=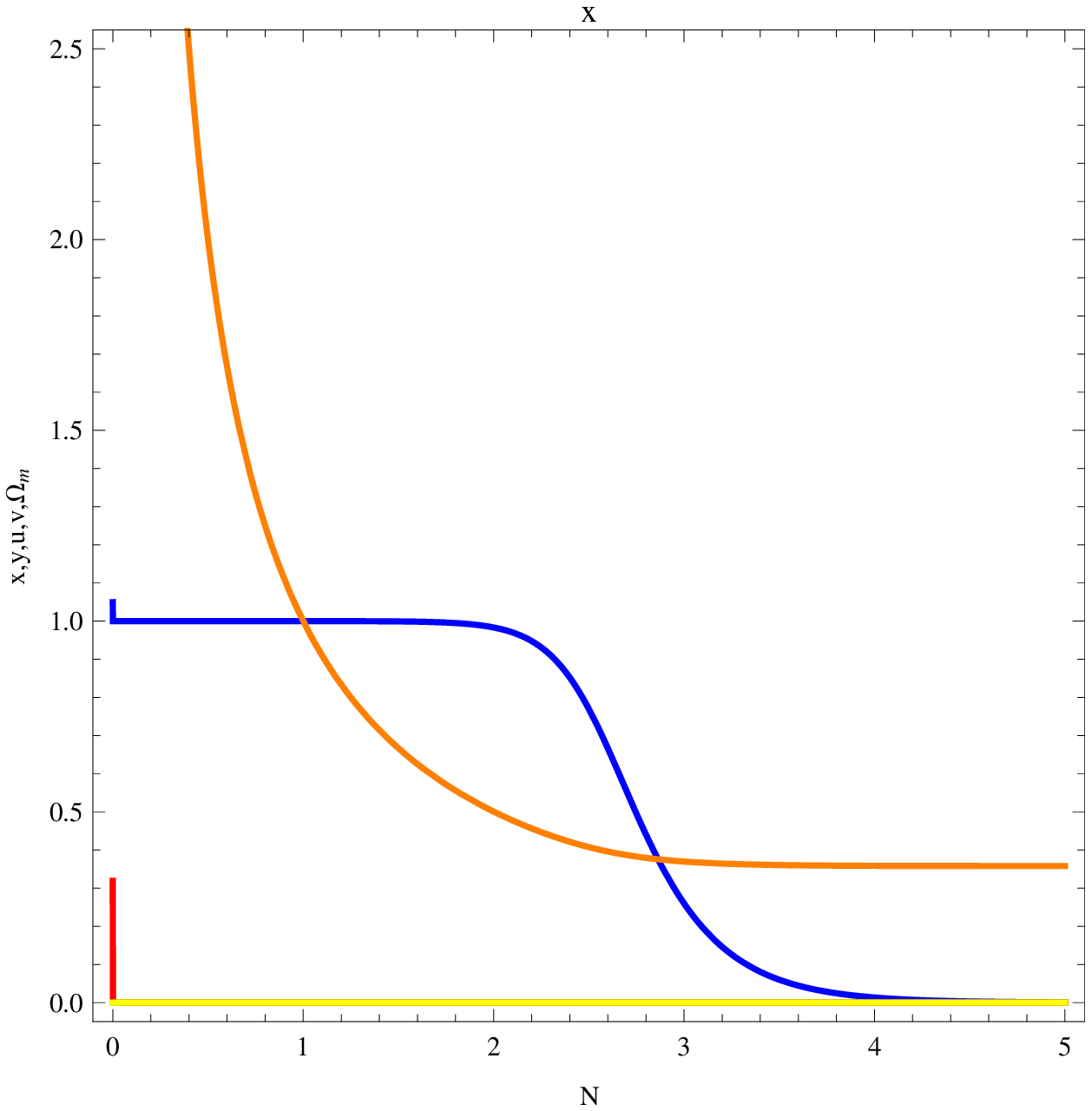,width=.50\linewidth}
\caption{Plot of (12) variations of x (blue),y (green),u
(orange),v (red),$\Omega_m$ (yellow) versus N near $P_{13}$,for $\gamma=1$ , $\delta=1$ and $\lambda=-0.5.$ The position corresponding N=0 is the fixed point
under consideration.}
\end{figure}
\begin{figure}\centering
\epsfig{file=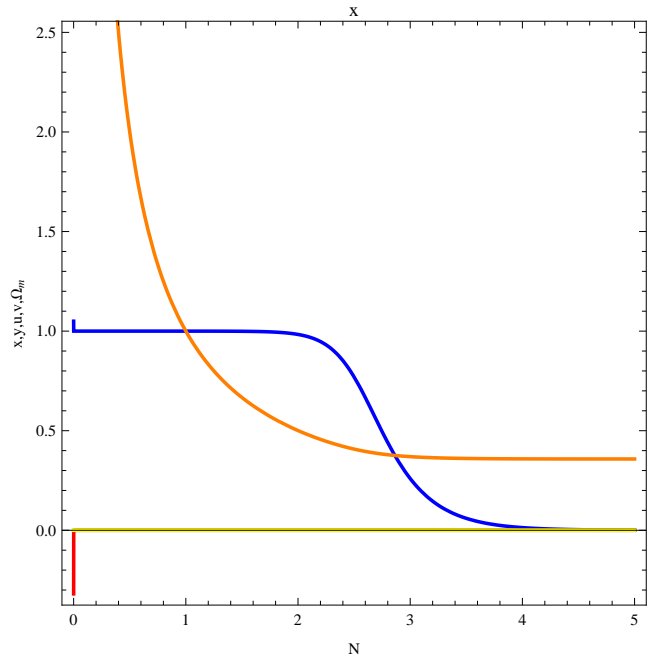,width=.50\linewidth}
\caption{Plot of (13) variations of x (blue),y (green),u (orange),v (red),$\Omega_m$
(yellow) versus N near $P_{14}$,for $\gamma=1$ , $\delta=1$ and
$\lambda=-0.5.$ The position corresponding N=0 is the fixed point
under consideration.}
\end{figure}
We note that if $2\delta<\lambda<0$ or $0<\lambda<2\delta$, then
$P_{13}$ and $P_{14}$ may admit 1 dimensional stable manifold corresponding
the negative eigenvalues with Eos of hessence and total Eos being 1 and universe decelerates.
The graphs figure $\mathbf{12}$ and figure $\mathbf{13}$ shows that the
system is diverging from the fixed point $P_{13}$ and $P_{14}$. So, both the points are unstable in nature.\\

\begin{figure}\centering
\epsfig{file=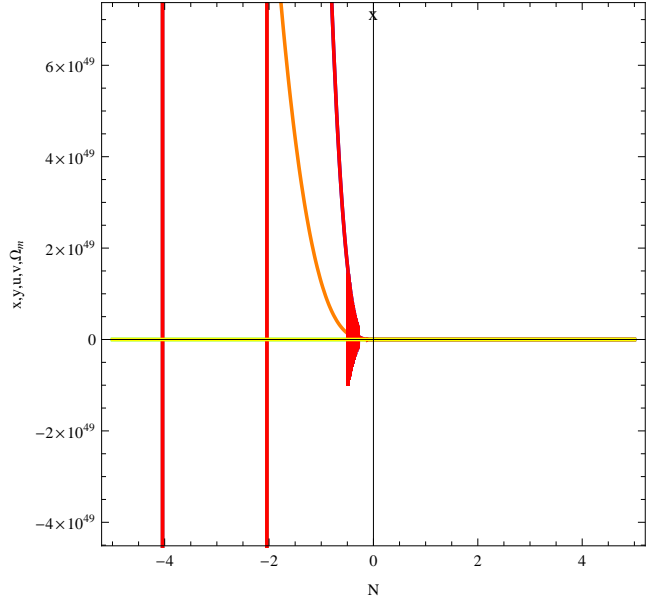,width=.50\linewidth}
\caption {Plot of (14) variations of x (blue),y (green),u
(orange),v (red),$\Omega_m$ (yellow) versus N near $P_{15}$,for $\gamma=1$ ,
$\delta=1/(4\sqrt{6})$ and $\lambda=-0.5.$ The position
corresponding N=0 is the fixed point under consideration.}
\end{figure}
\begin{figure}\centering
\epsfig{file=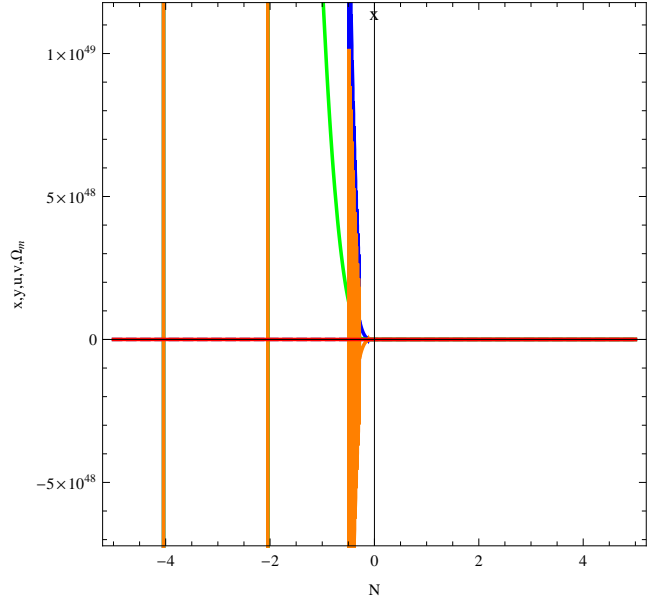,width=.50\linewidth}
\caption {Plot of (15) variations of x (blue),y (green),u (orange),v (red),$\Omega_m$
(yellow) versus N near $P_{16}$,for $\gamma=1$ ,
$\delta=1/(4\sqrt{6})$ and $\lambda=-0.5.$ The position
corresponding N=0 is the fixed point under consideration.}
\end{figure}
We note that if $x\delta{\leq}-\sqrt{\frac{3}{2}}$ and $x\lambda{\leq}-\sqrt{6}$
then $P_{15}$ and $P_{16}$ may admit 2 dimensional stable manifold corresponding
the two negative eigenvalues with Eos of hessence and total Eos being 1 and universe decelerates.
Here, we note that the solution set of the dynamical system moving rapidly
from the fixed points $P_{15}$ and $P_{16}$ as clear from figures $\mathbf{14}$
and $\mathbf{15}$.The fixed points are unstable.\\

\section{\textit{Cosmological Significance of the Fixed Points}}
In this section we discuss about the possible singularities that
any dark energy model could have and compare the fixed points
against recent dataset Planck 2015 data \cite{m8a-A1nn}. If the
Eos $\omega{\leq}-1$ (i.e., the null energy condition
$p+\rho{\geq}0$ is violated) and Big rip singularity happens
within a finite time \cite{7}. This singularity happens when at
finite time $t{\rightarrow}t_s$,
$a{\rightarrow}{\infty}$, $\rho{\rightarrow}{\infty}$ and $|p|{\rightarrow}{\infty}$.\\
We now analyse the stable fixed points to see if they can avoid (or, suffer)
Big rip singularity. For the stable fixed points $P_7$ and $P_8$ or, we have
$\dot{H}/{H^2}=0$ which gives $H=k$ (the integral constant), we get $a{\varpropto}{e^{kt}}$.
Also, in these cases ${\omega_{total}}=-1$ which with energy conservation equation
gives $\rho=constant$. Hence universe suffers no Big rip here. Fixed points $P_7$
and $P_8$  exist with physical parameter ${\Omega_m}=0$, ${\omega_h}=-1$, ${\omega_{total}}=-1$.
The value of the parameters are well within the best fit of Planck 2015 data i.e.,
${\Omega_m}=0.3089{\pm}0.0062$ from TT, TE, EE+low P+lensing+ext data, and Eos of
dark energy $\omega={-1.019}^{+0.075}_{-0.080}$ \\
Now, we consider the unstable fixed points. An unstable fixed
point may describe the initial phase of universe, whereas a stable
fixed point may be the end phase of the universe. For fixed points $P_1$, $P_2$ and $P_3$
exist with the physical parameter ${\Omega_m}=0$, ${\omega_h}=1$, ${\omega_{total}}=1$. Clearly,
no Big rip occurs here.Here, the parameter ${\Omega_m}$ lies within the best fit of Planck 2015
data i.e., ${\Omega_m}=0.3089{\pm}0.0062$ from TT, TE, EE+low P+lensing+ext data. But,${\omega_h}$,
${\omega_{total}}$ defy the  Eos of dark energy $\omega={-1.019}^{+0.075}_{-0.080}$.\\
Fixed point $P_4$ has values of physical parameters ${\Omega_m}=\frac{6-3\gamma+2\delta^2}{3}$,
${\omega_h}=1$, ${\omega_{total}}=-1 + \gamma(1-\frac{2{\delta^2}}{3})+ \frac{4{\delta^2}}{3}$.
Here, ${\omega_h}$ and  ${\omega_{total}}$ both are greater than -1, no Big rip occurs here too.
A wide choices of $\gamma$ and $\delta$ can can fit $\Omega_m$ and ${\omega_{total}}$ within
Planck 2015 data i.e., ${\Omega_m}=0.3089{\pm}0.0062$, but ${\omega_h}$, disobey the  Eos of
dark energy $\omega={-1.019}^{+0.075}_{-0.080}$. \\
Fixed points $P_5$ ,$P_6$  exist with physical parameters
${\Omega_m}=0$, ${\omega_h}=1$, ${\omega_{total}}=1$. We observe
this solution are devoid of Big rip.Here,${\Omega_m}$ lies within
the best fit of Planck 2015 data data. But,${\omega_h}$,
${\omega_{total}}$ defy the Eos of dark energy $\omega={-1.019}^{+0.075}_{-0.080}$.\\
Fixed points $P_9,P_{10}$ admit physical parameters as
${\Omega_m}=0$, ${\omega_h}=-1+\frac{\lambda^2}{3}$,
${\omega_{total}}=-1+\frac{\lambda^2}{3}$ and so avoid Big
rip.Also, $\Omega_m$ is within  Planck 2015 data. Also, suitable
choice of $\lambda$ fits ${\omega_h}$, ${\omega_{total}}$ within dataset.\\
Fixed points $P_{11}$ and $P_{12}$  have physical parameters
${\Omega_m}=B$, ${\omega_h}=\frac{-1+2{A^2}+{B^2}}{1-B}$,
${\omega_{total}}=-1+{A^2}+{B^2}+(\gamma-2)B$, where
$A=-\sqrt{\frac{3}{2}}~\frac{\gamma}{\delta+\lambda}$ and
$B=\frac{6+\lambda \sqrt{6}A-6{A^2}}{9}$. Here, we can adjust $A$
and $B$ to make ${\omega_h}$ and ${\omega_{total}}{\geq}-1$ to miss Big rip.
Since, only, $0<\gamma{\leq}2$ but, $\delta$ can take arbitrary small value
and $\lambda$ can have any real value, $A$ and hence, $B$ can be adjusted well
within Planck  ${\Omega_m}=0.3089{\pm}0.0062$ from TT, TE, EE+low P+lensing+ext
data and  Eos of dark energy $\omega={-1.019}^{+0.075}_{-0.080}$ data.\\
Fixed points $P_{13}$,$P_{14}$, $P_{15}$ and $P_{16}$ can avoid
Big rip, as they bear physical parameters ${\Omega_m}=0$,
${\omega_h}=1$, ${\omega_{total}}=1$.Here, the parameter
${\Omega_m}$ lies within the best fit of Planck 2015 data i.e.,
${\Omega_m}=0.3089{\pm}0.0062$ from TT, TE, EE+low P+lensing+ext
data. But, ${\omega_h}$, ${\omega_{total}}$ totally defy
the Eos of dark energy $\omega={-1.019}^{+0.075}_{-0.080}$.\\

\section{\textit{Concluding Remarks}}

In this paper we have performed a dynamical system study of an
unique scalar field hessence coupling with dark matter in an
alternate theory of gravity, namely $f(T)$ gravity. The system is
unconventional, complex but quite interesting. The model is chosen
to explore one of the various possibilities about the fate of
the universe. The sole purpose is to explain the current
acceleration of universe. An unstable fixed point may describe the
initial phase of universe, whereas a stable fixed point may be the
end of the universe. We have chosen exponential form of potential
of the form $V={V_0}{e^{\lambda\phi}}$ (where $V_0$ and $\lambda$
are real constant and $\phi$ is the hessence field) for
simplicity. The interaction term $C$ is chosen to solve the so
called `cosmological constant' problem in tune with second law of
thermodynamics and is quite arbitrary (only $C$ should remain
positive), since $C=\delta \dot{\phi} \rho_m$, where $\delta$ is a
real constant of small magnitude, which may be chosen as positive
or negative, such that $C$ remains positive. Also, $\dot{\phi}$
may be positive or negative according the hessence field $\phi$.
The resulting non linear dynamical system gives sixteen possible
fixed points. Among them $P_7$ and $P_8$ are stable set of
normally hyperbolic fixed points, which resembles like
`cosmological constant', so it explain the current phase of
acceleration of universe. But, interestingly it does not show
`hessence like' nature. Among the other fixed points the initial
phases of evolution may begin. However, the complexity of the
system is main obstacle for a precise explanation. Anyway, in
future work, we may try some other possible alternative.\\\\

{\bf Conflict of Interest:} The authors declare that there is no
conflict of interest regarding the publication of this paper.\\\\

{\bf Acknowledgement:} One of the authors (UD) is thankful to
IUCAA, Pune, India for warm hospitality where part of the work was
carried out.

\end{document}